\documentclass[useAMS]{mn2e}

\title[Infrared studies of Nova Scorpii 2014]{Infrared studies of Nova Scorpii 2014: an outburst in a symbiotic system sans an accompanying blast wave }

\author[Joshi et al.]{Vishal Joshi$^{1}$ \thanks{E-mail: vishal@iucaa.ernet.in},
D. P. K. Banerjee$^{2}$ \thanks{E-mail: orion@prl.res.in}, N M Ashok$^{2}$, V Venkataraman$^{2}$
\and $\&$ F.M. Walter$^{3}$\\
\\
$^{1}$Inter-University Centre for Astronomy and Astrophysics, Post Bag 4, Ganeshkhind, Pune University Campus, Pune 411007, India.\\
$^{2}$Astronomy \& Astrophysics Division, Physical Research Laboratory, Navrangpura, Ahmedabad 380009, India.\\
$^{3}$Department of Physics and Astronomy, Stony Brook University, Stony Brook, NY 11794-3800.\\}

\usepackage{graphicx}
\usepackage{array}

\begin{document}

\maketitle

\label{firstpage}

\begin{abstract}
Near-IR spectroscopy is presented for Nova Scorpii 2014. It is shown that the outburst occurred in a symbiotic binary
system - an extremely rare configuration for a classical nova outburst to occur in but appropriate for the eruption of
a recurrent nova of the T CrB class. We estimate the spectral class of secondary as M5III $\pm$ (two sub-classes). The maximum magnitude
versus rate of decline (MMRD) relations give an unacceptably large value of 37.5 kpc for the distance. The spectra are
typical of the He/N class of novae with strong HeI and H lines. The profiles are broad and flat topped with full width
at zero intensities (FWZIs) approaching 9000-10000 km s$^{-1}$ and also have a sharp narrow component superposed which
is attributable to emission from the giant's wind. Hot shocked gas, accompanied by X-rays and $\gamma$ rays, is expected
to form when the high velocity ejecta from the nova plows into the surrounding giant wind. Although X-ray emission was
observed no $\gamma$-ray emission was reported. It is also puzzling that no signature of a decelerating shock is seen
in the near-infrared (NIR), seen in similar systems like RS Oph, V745 Sco and V407 Cyg, as rapid narrowing of the line
profiles. The small outburst amplitude and the giant secondary strongly suggest that Nova Sco 2014 could be a recurrent nova.

\end{abstract}



\section{Introduction}

Nova Scorpii 2014 was first discovered as a transient by Nishiyama and Kabashima (2014) on 2014 March 26.84867 UT
at a magnitude of 10.1 in unfiltered CCD images. The discovery was announced in Special Notice $\#$383 of
the American Association of Variable Star Observers (AAVSO) which designated the source as TCP J17154683-3128303.
Survey frames, by the observers above, from 2014 Mar 22.819 UT (limiting magnitude of 12.5) and 23.836 UT
(limiting magnitude of 12.9) showed no object at the transients's position, nor did DSS or USNO-B1.0 plates.
The object was confirmed to be a nova by Ayani and Maeno (2014) on 2014 March 27.8 UT through a low-resolution
spectrogram which showed broad emission lines of Balmer series, He\,{\sc i} 501.6, 587.8, 706.5, and probably O\,{\sc i} 777.3 nm.
The H$\alpha$ line had an full width at half maximum (FWHM) of 7000 km s$^{-1}$ and EW of about 90 nm. A reconfirmation of the object as a nova
was obtained by Jelinek et al. (2014) through an an optical spectrum taken on 2014 March 30.19 UT. The rapid
photometric decline of the object was subsequently monitored by Munari et al. (2014) who noted that its large $B-V$
color suggested a large reddening affecting the nova.

\begin{table*}
\begin{center}
\begin{minipage}{138mm}
\caption{Log of the spectroscopic observations}

\newcolumntype{L}[1]{>{\raggedright\let\newline\\\arraybackslash\hspace{0pt}}m{#1}}
\newcolumntype{C}[1]{>{\centering\let\newline\\\arraybackslash\hspace{0pt}}m{#1}}
\newcolumntype{R}[1]{>{\raggedleft\let\newline\\\arraybackslash\hspace{0pt}}m{#1}}

\begin{tabular}{ C{3cm} C{2cm} C{1.5cm} C{0.5cm} C{1.5cm} C{1.0cm} C{1.2cm} }
\hline
Date of          & Days since    &\multicolumn{3}{c}{Total exposure time}     & \multicolumn{2}{c}{Airmass} \\
Observation      & outburst$^a$  &             &           &        &         &                   \\
(UT)             & (days)        &\multicolumn{3}{c}{(seconds)}     &         &                   \\
                 &               &  $IJ$       & $JH$      & $HK$   & Nova    & Standard          \\
\hline
2014 March 30.98 &  4.13         &  720        & 720       & 1440   & 1.82    & 1.75              \\
2014 April 02.98 &  7.12         & 1260        & 300       &  600   & 1.81    & 1.74              \\
2014 April 03.99 &  8.13         &  600        & 300       & 1080   & 1.80    & 1.74              \\
2014 April 05.99 & 10.13         &  760        & ---       &  760   & 1.80    & 1.75              \\
2014 April 09.01 & 13.13         &  760        & 760       &  760   & 1.81    & 1.74              \\
2014 April 15.01 & 19.14         &  760        & 570       &  760   & 1.86    & 1.79              \\
\hline
\end{tabular}
\scriptsize{$^a$ Time of outburst is taken as 2014 March 26.84867 UT or JD 2456743.34867 (Nishiyama \& Kabashima, 2014)}
\end{minipage}
\end{center}
\end{table*}

Near-infrared observations by Joshi et al (2014) obtained on 2014 March 30.98 UT in the 0.85-2.4 $\mu$m region showed
a spectrum typical of the He/N class of novae dominated by broad hydrogen lines of the Paschen and Brackett series
and strong He\,{\sc i} lines at 1.083 and and 2.058 $\mu$m. However, the most striking feature was the presence of first
overtone absorption bands of CO at 2.2935 $\mu$m and beyond suggesting that Nova Sco 2014 was a symbiotic system
with a cool giant secondary. The FWZI's of the Paschen $\beta$, Brackett $\gamma$, He\,{\sc i} (2.058 $\mu$m) and O\,{\sc i} (1.1287 $\mu$m)
lines were reported as 8900, 9500, 10100, 8800 km s$^{-1}$ respectively. Photometry on 2014 March 29 showed the source to
have NIR magnitudes of $J$ = 8.80 $\pm$ 0.03, $H$ = 8.31 $\pm$ 0.04 and $K$ = 7.62 $\pm$ 0.02.

Interestingly, the object was detected in X-rays by Swift almost concurrent with the optical discovery announcement
(Kuulkers et al 2014). Two Swift snapshots, taken 7.3 and 8.4 hours after the discovery (i.e., on 2014 March 27 03:39-03:51
and 04:42-04:51 UT), revealed a new, bright, X-ray source that was also seen in UVOT. The mean X-ray spectrum was best fit
with an absorbed optically thin emission model with most of the absorption being intrinsic to the source and consistent with
an (expanding) shell in a nova. Continued Swift observations of the nova were also reported by Page et al (2014). The details
and implications of the X-ray observations on our results are discussed later.

In this paper we present detailed NIR spectroscopic observations of Nova Sco 2014. The realization that this nova system
contained a cool giant secondary, considerably enhanced our interest because strong shocks are known to be generated in
such symbiotic systems when the nova ejecta collides with the surrounding giant wind. The shocked gas can become the site
of X-ray and $\gamma$-ray generation - it may be noted that $\gamma$-ray detections from novae is a developing field of
considerable interest. However no $\gamma$-ray emission was detected from this nova. We discuss this, and other aspects,
of this interesting and rather unique nova in the coming sections.

\section{Observations}
Near-IR spectroscopy in the 0.85 to 2.4 $\mu$m region at resolution $R$ $\sim$ 1000 was carried out with the 1.2m telescope
of the Mount Abu Infrared Observatory using the Near-Infrared Camera/Spectrograph (NICS) equipped with a
1024x1024 HgCdTe Hawaii array. The instrument has an 8$\times$8 arc minute square unvignetted field of view.
Spectra were recorded with the star dithered to two positions along the slit with one or more spectra being
recorded in both of these positions. The coadded spectra in the respective dithered positions were subtracted
from each other to remove sky and dark contributions. The spectra from these sky-subtracted images were
extracted using IRAF\footnote{ IRAF is distributed by the National Optical Astronomy Observatory, which
is operated by the Association of Universities for Research in Astronomy (AURA) under a cooperative
agreement with the National Science Foundation.} tasks and wavelength calibrated using a combination of OH sky lines and telluric lines
that register with the stellar spectra. To remove telluric lines from the target's spectra, it was ratioed
with the spectra of a standard star (SAO 208606 , spectral type B9.5V, T$_{eff}$ = 10,000K) from whose spectra
the hydrogen Paschen and Brackett absorption lines had been removed. The spectra were finally multiplied by
a blackbody at the effective temperature of the standard star to yield the resultant spectra. All spectra were covered
in three settings of the grating that cover the $IJ, JH$ and $HK$ regions separately. The log of the
spectroscopic observations is given in Table 1.

\begin{figure}
\centering
\includegraphics[bb= 0 0 311 329,width=3.25in,height= 3.25in,clip]{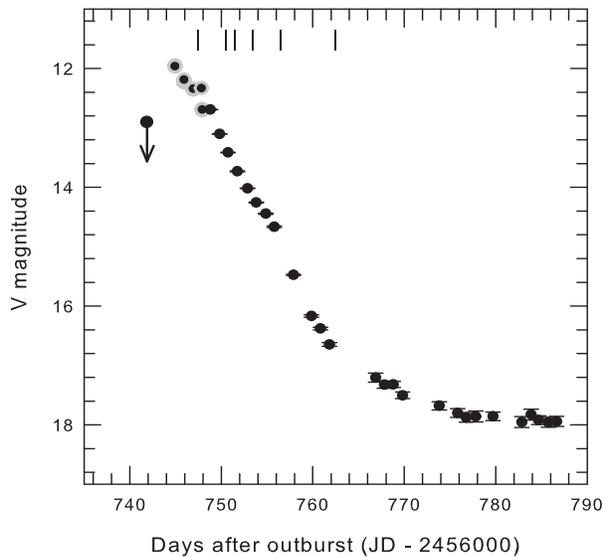}
\caption[]{ The observed V band light curve of Nova Sco 2014 {\bf uncorrected for reddening}. Gray points are from AAVSO, black filled circles are from SMARTS data.
Error bars are included. Epochs of our spectroscopic observations are marked by vertical ticks on top. The outburst
date is taken as 2014 March 26.84867 UT equal to JD 2456743.34867. The nova was not detected up to a limiting magnitude
of 12.9 just 3 days before outburst (arrow mark) indicating it was caught close to maximum light. }
\label{fig3}
\end{figure}

\section{Results}

\subsection{Light curve and Extinction}
Figure 1 shows the $V$ band light curve of the nova using data taken from the SMARTS observations
(Walter et al. 2012) except for the first 3 to 4 days where we have included AAVSO data to extend
coverage towards maximum. The non-detection of the nova, just 3 days before discovery indicates the
nova was caught close to maximum light. From the AAVSO data we determine t$_{2}$ - the time for the brightness
to decline by 2 magnitudes from maxima - to be 6 $\pm$ 0.3d thereby putting it in the \textbf{very} fast speed class.

The observed $(B-V)$ values near maximum and at t$_{2}$ taken from the AAVSO database are equal to 1.00 and 1.03
respectively in contrast to the generally expected values of 0.23 $\pm$ 0.06 and -0.02 $\pm$ 0.04
respectively at these epochs (Van den bergh and Younger, 1987). The large values of $(B-V)$ imply
considerable reddening; the excess $E(B-V)$ values are equal to 0.77 and 1.05 respectively.
We adopt a mean value of 0.91 for the reddening and thus an extinction
A$_{V}$ = 3.09$\times E(B-V)$ = 2.81 (Rieke \& Lebofsky (1985)). Schlafly and Finkbeiner (2011) give a mean value of
$E(B-V)$ = 0.93 and A$_{V}$ = 2.883 in the direction of the nova from dust extinction maps. For t$_{2}$ = 6d,
the MMRD relation of della Valle and Livio (1995) gives M$_{V}$ = -8.87 which implies an extremely large distance
to the nova of 37.5 kpc. This places the nova outside the Milky Way which does not appear plausible. The
application of MMRD relationships clearly gives a discrepant result for this particular nova. A similar
problem occurs quite often when applied to the recurrent novae (Schaefer 2010).

\begin{figure}
\centering
\includegraphics[bb= 49 89 602 702, width=3.25in,height= 3.25in,clip]{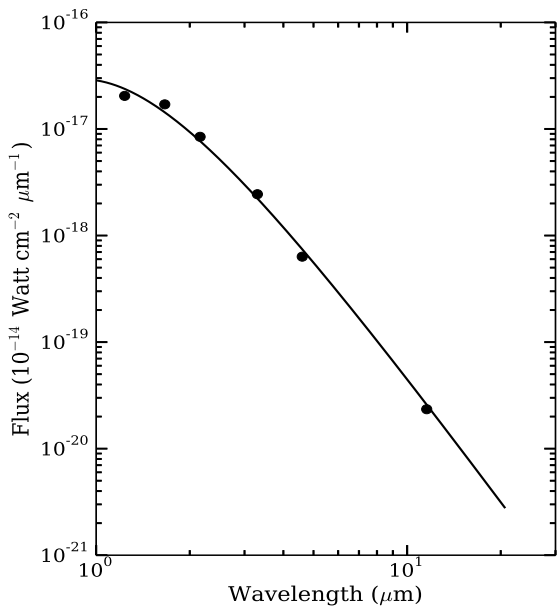}
\caption[]{The SED of Nova Sco 2014 in quiescence plotted using dereddened 2MASS and WISE data and
fitted by a blackbody at 3200 $\pm$ 50K.}
\label{fig3}
\end{figure}

\begin{table}
\begin{center}
\caption{A list of the lines identified from the $JHK$ spectra.}
\begin{tabular}{lr}
\hline\\
Wavelength     & Species     \\
(${\mu}$m)     &             \\
\hline                       \\
0.9229         & Pa 9        \\
0.9546         & Pa 8        \\
1.0049         & Pa 7        \\
1.0124         & He\,{\sc ii}\\
1.0830         & He\,{\sc i} \\
1.0938         & Pa $\gamma$ \\
1.1287         & O\,{\sc i}  \\
1.2461, 1.2469 & N\,{\sc i}  \\
1.2818         & Pa $\beta$  \\
1.5256         & Br 19       \\
1.5341         & Br 18       \\
1.5439         & Br 17       \\
1.5557         & Br 16       \\
1.5701         & Br 15       \\
1.5881         & Br 14       \\
1.6109         & Br 13       \\
1.6407         & Br 12       \\
1.6806         & Br 11       \\
1.7002         & He\,{\sc i} \\
1.7362         & Br 10       \\
1.9451         & Br 8        \\
2.0581         & He\,{\sc i} \\
2.1120, 2.1132 & He\,{\sc i} \\
2.1656         & Br $\gamma$ \\
2.2935, 2.3227 & CO 2-0, 3-1 and    \\
2.3535, 2.3829 & 4-2, 5-3 bandheads    \\
\hline
\end{tabular}
\end{center}
\end{table}

\begin{table*}
\begin{center}
\begin{minipage}{148mm}
\caption{A list of emission line flux values ($10^{-20}$ Watt cm$^{-2}$) at different epochs. The numbers in the paranthesis
indicate the associated error. }
\begin{tabular}{@{} l c c c c c c @{}}
\hline
  Emission                   &   \multicolumn{6}{c}{Line flux at Days after Outburst}  \\
    Line                     &  4.13         &  7.12         &  8.13        &  10.13        &  13.13       &  19.14       \\
\hline
  Pa 7+He\,{\sc ii} 1.0124    &  168.0 (25.0) &   59.2 (10.4) &  47.9 (9.4)  &  42.1 (8.2)   &  20.9 (4.8)  &  7.28 (1.85)  \\
  He\,{\sc i} 1.08+Pa $\gamma$ & 1460.0 (76.0) &  684.0 (69.0) & 584.0 (67.0) & 406.0 (64.72) & 243.0 (48.0) &  69.7 (15.4)  \\
  O\,{\sc i} 1.1287          &  219.0 (35.0) &   98.5 (16.6) &  58.8 (11.4) &  38.5 (7.6)   &  10.1 (2.4)  &  5.00 (1.20)  \\
  N\,{\sc i} 1.2461 + 1.2469 &   11.8 (2.1)  &   8.18 (1.41) &  5.71 (1.14) &  4.79 (0.96)  &  6.25 (1.35) &  2.37 (0.57)  \\
  Pa $\beta$                 &  187.0 (27.0) &   64.4 (11.4) &  52.3 (10.2) &  31.6 (6.2)   &  19.3 (4.0)  &  7.74 (1.74)  \\
  O\,{\sc i} 1.31            &   ---         &   5.77 (1.04) &  7.97 (1.59) &  2.80 (0.56)  &  1.57 (0.41) &  ---          \\
  Br 19                      &   1.57 (0.25) &   1.44 (0.24) &  ---         &  ---          &  ---         &  ---          \\
  Br 18                      &   1.79 (0.30) &   1.54 (0.26) &  ---         &  ---          &  ---         &  ---          \\
  Br 17                      &   3.29 (0.61) &   2.99 (0.51) &  2.46 (0.49) &  ---          &  ---         &  ---          \\
  Br 16                      &   5.50 (1.0)  &   4.03 (0.70) &  7.48 (1.49) &  ---          &  ---         &  ---          \\
  Br 15                      &   6.35 (1.21) &   5.57 (1.05) &  5.51 (1.10) &  ---          &  ---         &  ---          \\
  Br 14                      &   9.48 (1.82) &   6.27 (1.21) &  7.81 (1.56) &  ---          &  3.77 (0.95) &  ---          \\
  Br 13                      &   10.9 (2.0)  &   6.27 (1.20) &  6.45 (1.29) &  ---          &  4.31 (0.96) &  ---          \\
  Br 12                      &   17.3 (3.1)  &   11.5 (2.1)  &  14.4 (2.9)  &  6.12 (1.22)  &  11.2 (2.4)  &  ---          \\
  Br 11 + He\,{\sc i} 1.7002 &   26.2 (4.5)  &   14.6 (2.5)  &  16.9 (3.4)  &  10.3 (2.0)   &  8.96 (1.98) &  ---          \\
  Br 10                      &   20.4 (3.6)  &   12.3 (2.2)  &  10.1 (2.0)  &  11.2 (2.2)   &  5.64 (1.22) &  ---          \\
  He\,{\sc i} 2.0581         &   89.2 (15.0) &   57.6 (10.1) &  42.8 (8.4)  &  27.5 (5.4)   &  11.0 (2.3)  &  ---          \\
  Br $\gamma$                &   30.7 (4.0)  &   10.4 (1.5)  &  7.61 (1.52) &  7.09 (1.61)  &  3.20 (0.7)  &  ---          \\

\hline
\end{tabular}
$^a$ Time of outburst is taken as 2014 March 26.84867 UT or JD 2456743.34867 (Nishiyama \& Kabashima, 2014)
\end{minipage}
\end{center}
\end{table*}

\begin{figure}
\centering
\includegraphics[bb= 169 54 442 738,width=3.0in,height= 7.2in,clip]{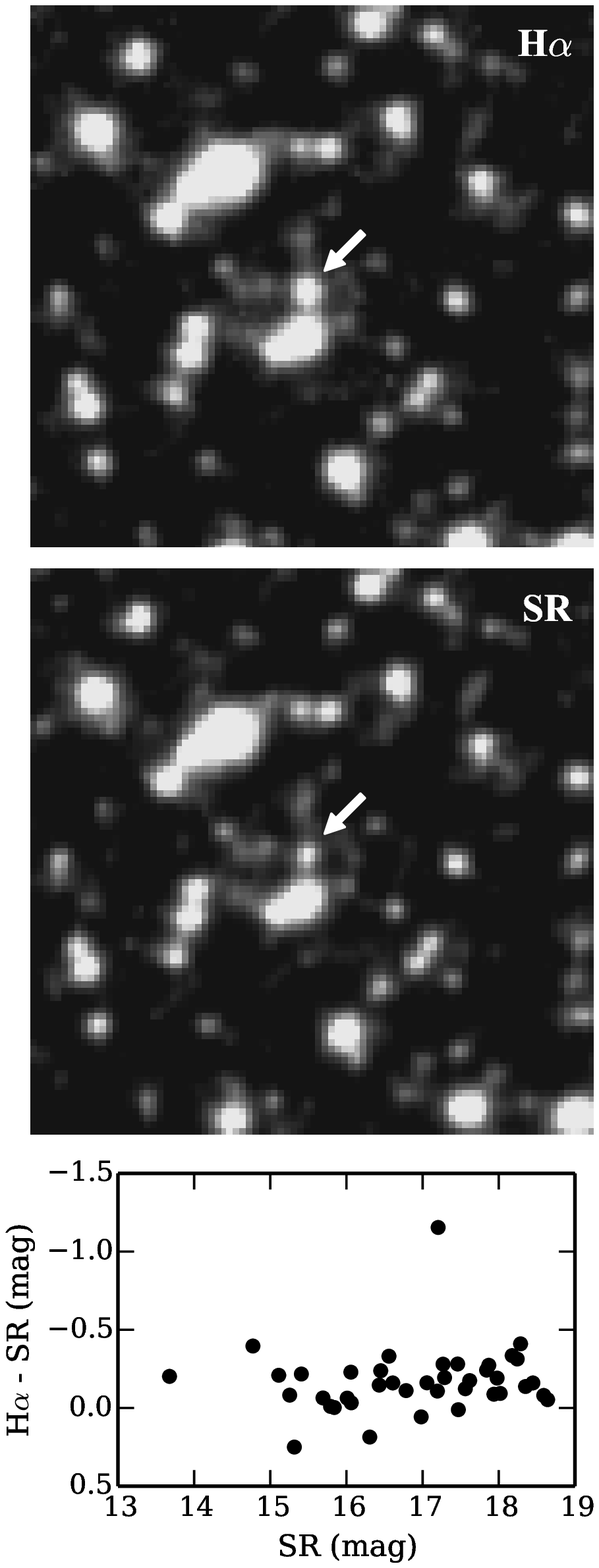}
\caption[]{The {\bf 2x2 arcminute square} H$\alpha$ and short $R$ (SR) band images from the SUPERCOSMOS database with the position of the nova marked.
The bottom panel shows that Nova Sco 2014 is a prominent H$\alpha$ emitter with a large (H$\alpha$ - $R$) excess of approximately
1.2 magnitudes (the isolated point at the top of the plot)}
\label{fig3}
\end{figure}

\subsection{Distance to the nova and nature of the secondary}

In view of the above, we therefore estimate the distance through an alternative approach by considering the
properties of the secondary. One of its striking properties is that it displays prominent first overtone
CO bands in the $K$ band at 2.2935 $\mu$m and beyond (this is shown in Figure 6 and discussed further in section 3.5).
The presence of these CO bands conventionally indicates that the secondary is a cool, late type star of
M type. Confirmation of the M spectral class is seen from the SED of the system in quiescence shown in
Figure 2. Herein we have used only longer wavelength data from 2MASS and WISE so that errors arising from
reddening corrections are minimized or negligible. Values of the 2MASS magnitudes used are $J$ = 11.26,
$H$ = 10.05 and $K$ = 9.58. These magnitudes are transformed to common filter system using transformation
equations given by Carpenter (2001) and corrected for extinction using relations given by Rieke \& Lebofsky (1985).
We have also used WISE magnitudes W$_{1}$ (3.3 $\mu$m)= 8.79, W$_{2}$ (4.6 $\mu$m) = 8.91 and
W$_{3}$ (11.56 $\mu$m) = 8.65. The SED is well fit by a black body of around 3200 $\pm$ 50 K
suggesting a spectral class of M6 ($\pm$ one sub-class) if the secondary is a giant.
We presume the secondary is a giant and not a main sequence star for the following reasons.
The corrected 2MASS magnitudes yield intrinsic $(J - H)$ and $(H - K)$ colors of 0.97
and 0.27 respectively. These colors are consistent with values expected for a M5/M6 giant viz. 0.96 and 0.28 respectively (Bessell and Brett; 1988, Table 3 therein). On the other hand a dwarf of similar spectral class would
be expected to have corresponding values of $\sim$ 0.66 and $\sim$ 0.37 respectively (Bessell and Brett; 1988, Table 2
therein; Straizys and Lazauskaite, 2009; Table 1 therein). Further the dereddened $(J - H)$ and $(H - K)$ colors
are typical of bulge giants and giant secondaries of symbiotic
stars in the $(J - H)$ versus $(H - K)$ color-color diagram of Whitelock and Munari (1992; see Figure 6 therein).
It would thus appear that the secondary in the system is a giant rather than a main sequence star. This would make
the Nova Sco 2014 system a symbiotic system, fairly rare among novae composing of a WD and a late type
giant secondary similar to a few other well known novae like V407 Cyg, RS Oph and V745 Sco. The SED in
Figure 2 does not show any IR excess due to dust indicating that this is a S type (stellar) rather than a
D type (dusty) system. D type systems generally contain a Mira variable and their near-IR colors
indicate the combination of a reddened Mira and dust with temperature of approximately 1000 K
(Belczynski et al. 2000).

We may add another argument favoring a giant rather than a dwarf status for the secondary.
For the main sequence dwarf stars the temperature value of 3200 $\pm$ 50 K derived earlier corresponds
to a spectral type of M4V and has the following physical parameters, namely, (V-K) = 5.250, M$_{V}$ = 12.80
and M$_{K}$=7.55 (Pecaut and Mamajek 2013; see online version of Table 5 therein).
The distance modulus relation then yields an unusually small value of 22$\pm$10 pc to
the object. However, such a close by object may be expected to have very little extinction vis-a-vis that which is observed. The
extinction models of Marshall et al (2006), in a field of radius 6 arc minutes centered on the source, have
A$_{K}$ values of 0, 0.204$\pm$0.051, 0.300$\pm$0.040 and 0.300$\pm$0.033 at distances of 0, 1.52, 3.13 and 8.95 kpc respectively.
A linear interpolation of this data between 0 to 1.52 kpc would suggest that at 22 pc the extinction A$_{K}$ should be 0.002 leading to A${_V}$=0.02
instead of the observed value of A${_V}$ = 2.81.

From the general equation $m - M = 5logd - 5 + A$ we make an attempt to
estimate the distance assuming a giant class for the secondary component
of Nova Sco 2014. The distance determination has a strong dependence on the
spectral subclass of the M giant secondary and we consider an error of two
subclasses from M5III. For M5III star intrinsic color $(V-K)_{intr}$ = 5.96, M$_V$ $\sim$ -0.3 and we
thus get M$_K$ = -6.26. In quiescence, the secondary will totally dominate
all contribution to the K band and thus m$_k$ will equal the quiescent 2MASS
magnitude. From Rieke and Lebofsky (1985) we  use A$_K$ = 0.112A$_V$ = 0.31 and using the known values of  m$_k$ and M$_K$, the distance from the distance-modulus relation is found to be 13 kpc. A similar calculation for other spectral classes
yields distance estimates of 8.1, 9.6, 13.0, 18.6 and 26.4 kpc for M3III, M4III, M5III, M6III and M7III spectral classes respectively.
Thus the distance estimate is very sensitive to a proper determination of the spectral class and prone to error.
At best we can say, assuming that a spectral sub-class earlier than M3III is unlikely,
 the lower limit to the distance to the object is 8 kpc. In all the above calculations we have used
 $(V-K)_{intr}$ from  Bessel and Brett (1988) which lists these values {\bf up to} M7III. The   M$_V$
 values have been taken from Lang (1990) for which M$_V$ values are listed upto M5III. For
 M6III and M7III we have extrapolated Lang's values and used M$_V$ = -0.2 and -0.1 for M6III and M7III
 respectively.


\begin{figure}
\centering
\includegraphics[bb=100 110 490 662,width=3.5in,height=5.0in,clip]{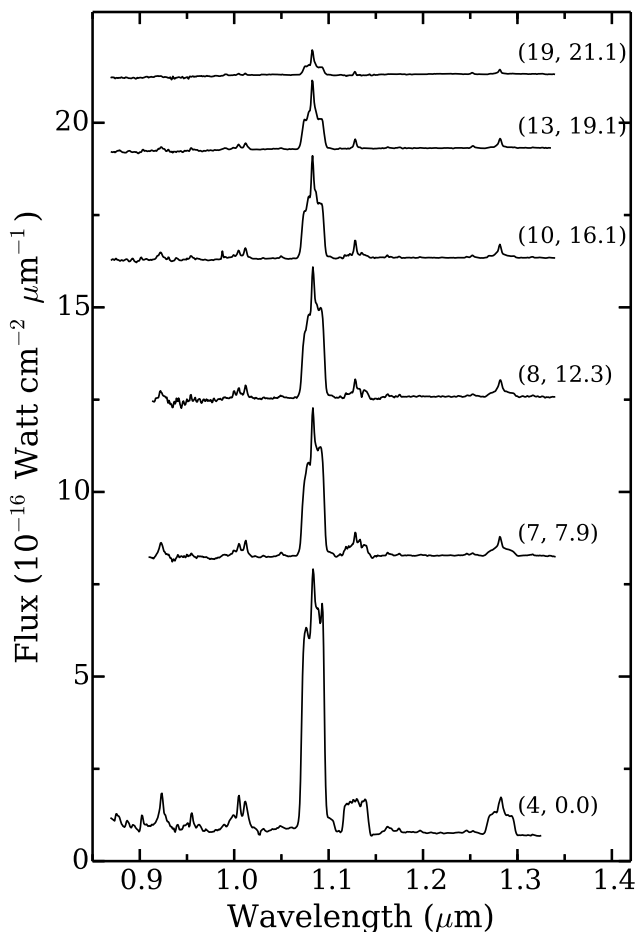}
\caption[]{ The $J$ band spectra of Nova Sco 2014. The numbers in the parentheses indicates the days after outburst and
offsets applied, for sake of clarity,  to the flux respectively. The emission lines are identified in Table 2.}
\label{fig3}
\end{figure}

\begin{figure}
\centering
\includegraphics[bb=100 110 490 662,width=3.5in,height=5.0in,clip]{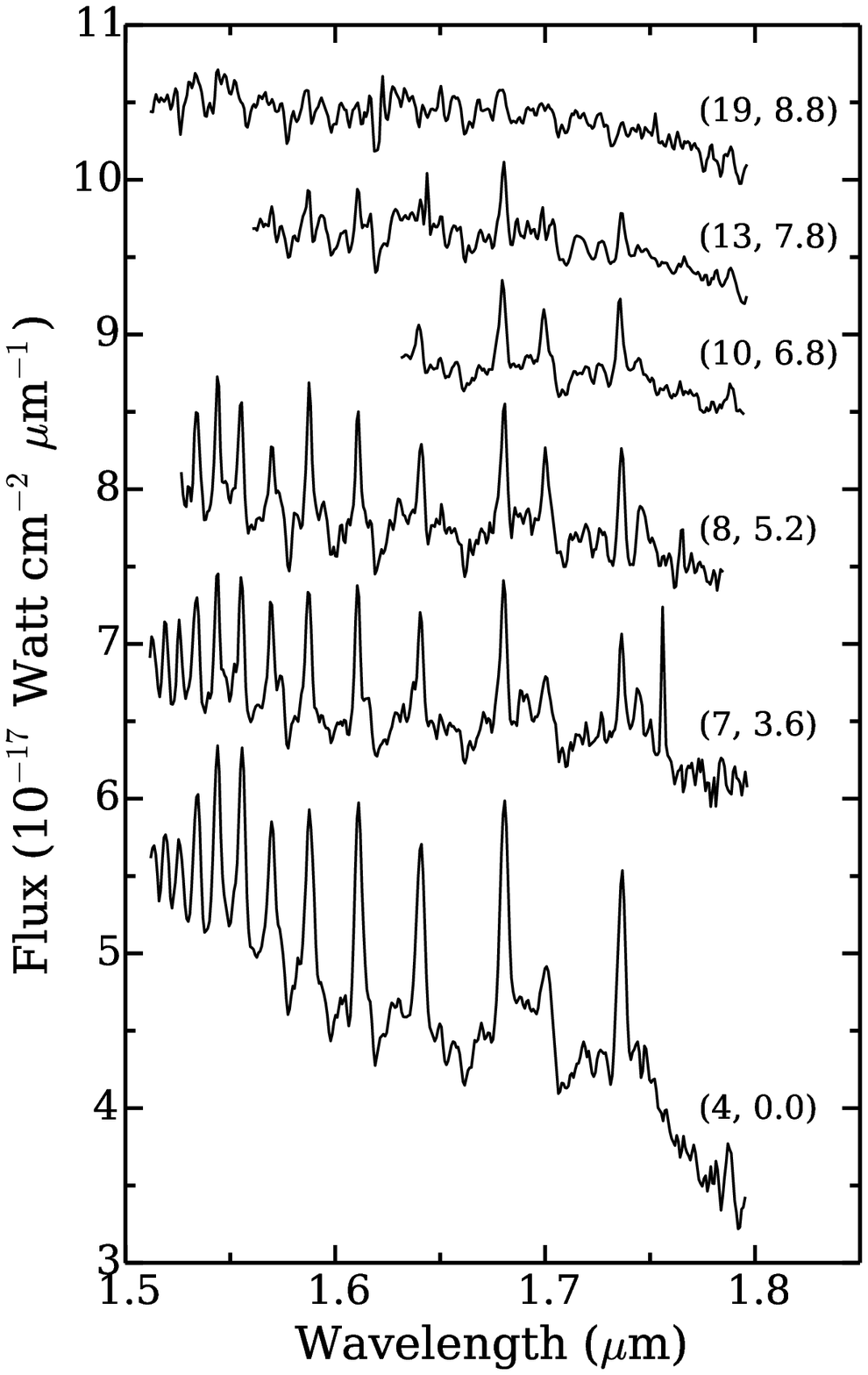}
\caption[]{ The $H$ band spectra of Nova Sco 2014. The numbers in the parentheses indicates the days after outburst and
offsets applied, for sake of clarity,  to the flux respectively. The emission lines are identified in Table 2.}
\label{fig3}
\end{figure}

\begin{figure}
\centering
\includegraphics[bb=100 110 490 662,width=3.5in,height=5.0in,clip]{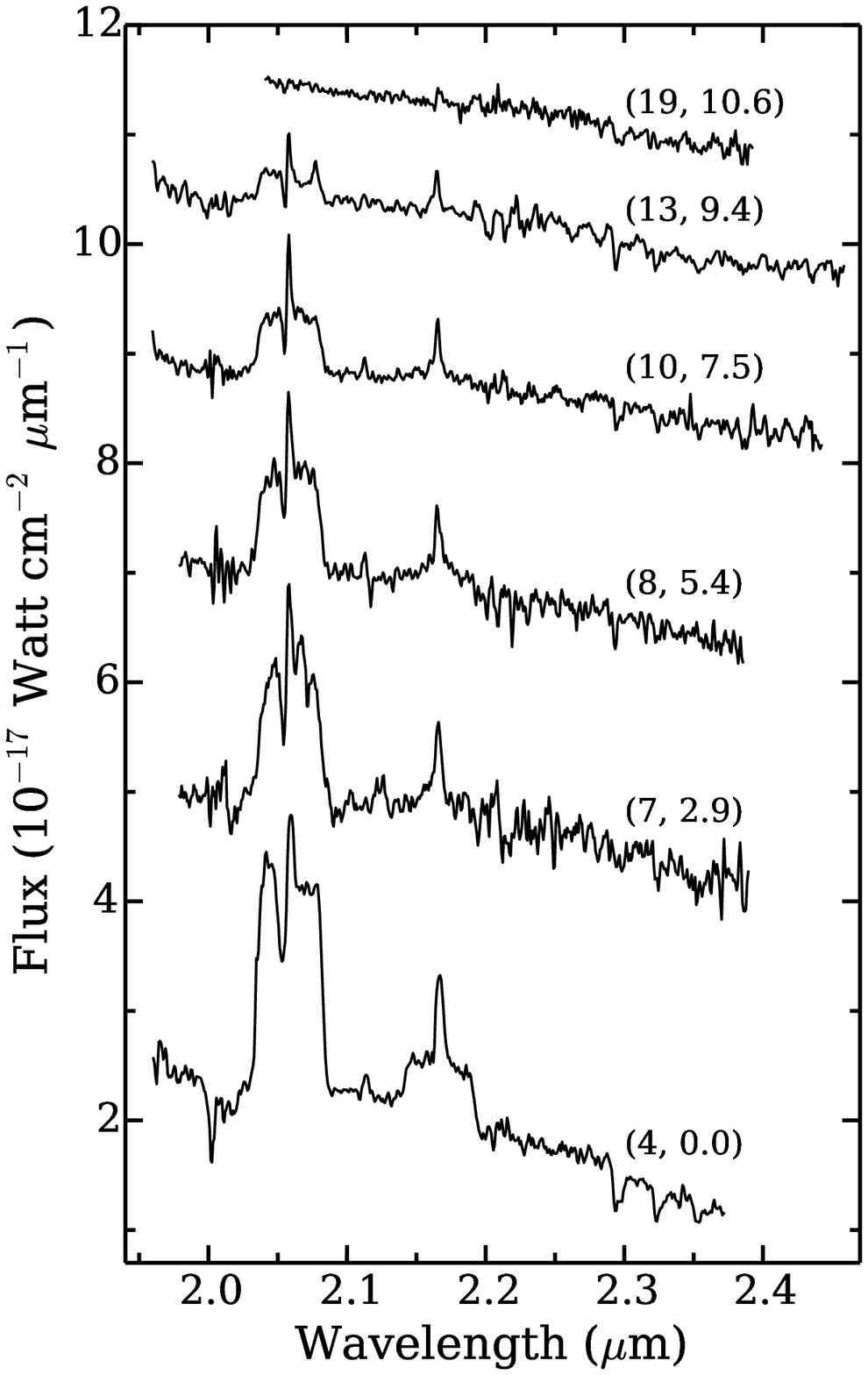}
\caption[]{ The $K$ band spectra of Nova Sco 2014. The numbers in the parentheses indicates the days after outburst and
offsets applied, for sake of clarity,  to the flux respectively. The emission lines are identified in Table 2.}
\label{fig3}
\end{figure}

\subsection{Outburst amplitude}
The outburst amplitude of this nova is small compared to that expected for a classical nova (CN). The quiescent
$B$ magnitude is 20.175 from SUPERCOSMOS plates while from the AAVSO data, $B$ = 12.96 at maximum, yielding an
outburst amplitude of 7.2 in the $B$ band. Continued monitoring of the object from SMARTS, up to 215 days after
outburst, shows that the object reaches a median quiescent magnitude of $V$ = 18.6 by 50d after outburst by which
time all flux from the system is essentially dominated by the secondary. However, there are small modulations in
brightness seen in the optical bands during this stage, the nature and cause of which is being studied and results of
which we {\bf will} present separately. Given that at peak brightness V$_{max}$ $\sim$ 12.0, an outburst amplitude of
$\sim$ 6.6 in $V$ is suggested.

The small amplitude of outburst in the $B$ and $V$ bands suggests the object could be a RN (discussed further in section 4).
To compare the outburst amplitudes of RN and CN it is necessary that the contribution from the secondaries
be subtracted in the case of the RNe. Without this subtraction or correction, all RN except T Pyx lie 2
to 5 magnitudes below the outburst amplitude (A) versus t$_{2}$ plot and this is an useful signature of
potential RN (Warner 1995; figure 5.4). This is true for this object; had it been a CN it is expected
that for t$_{2}$ = 6d, its amplitude should have been around 13 $\pm$ 1 magnitude instead of $\sim$ 6 to 7
observed here.\

\subsection{The symbiotic nature from H$\alpha$ images}
The symbiotic nature of this nova system is established more firmly by comparing its H$\alpha$ and R band
images shown in Figure 3, based on data available at the SUPERCOSMOS archive
(http://www-wfau.roe.ac.uk/sss/H$\alpha$/). The position of the nova is marked and relative to the other
stars, it is clearly seen to be considerably brighter in H$\alpha$ than in the $R$ band (short-red or SR)
image. The bottom panel shows a plot of the (H$\alpha$ - $R$) versus R magnitude of all stars in a field
2$\times$2 arc minute square centered around the nova. While most stars cluster in a narrow strip in this plot,
Nova Sco 2014 stands out with a large (H$\alpha$ - $R$) excess of approximately 1.2 magnitudes. Symbiotic stars
are known to be strong H$\alpha$ emitters (see spectra in the Munari and Zwitter 2002 catalog) and the
presence of such strong H\,{\sc i} emission is one of the key criteria for classifying them as symbiotic
stars (Belczynski et al. 2000).\

\subsection{General properties of the Spectra}
Due to the rapid decline in its brightness, spectra of the nova could be obtained only until 19d after
outburst; beyond which it was too faint to pursue. The flux calibration of
the spectra was done using either contemporaneous or interpolated values of
JHK magnitudes from the SMARTS database (http://www.astro.sunysb.edu/fwalter/SMARTS/
NovaAtlas/) matching the epochs of the Mt.Abu observations. These spectra, presented in Figures 4, 5 and 6, are
typical of the He/N class of novae and show strong helium lines apart from lines of hydrogen. Lines of
carbon, which are a distinguishing feature in the NIR of the Fe\,{\sc ii} class of nova, are totally absent.
The observed spectra can be compared with prototypical templates of the He/N class of novae in the NIR as
given in Banerjee and Ashok (2012). The prominent lines shown are the Brackett and Paschen recombination
lines of H\,{\sc i}, He\,{\sc i} lines at 1.0831, 1.7002 and 2.0581 $\mu$m with the 1.0831 $\mu$m line being
overwhelmingly strong, the N\,{\sc i} blend at 1.2461, 1.2469 $\mu$m. The O\,{\sc i} line at 1.1287 $\mu$m usually
seen in the novae is also present and the absence of O\,{\sc i} 1.3164 $\mu$m line indicates that the Lyman-$\beta$
fluorescence is the dominant emission process (Rudy et al 1991). The detailed list of observed lines is presented in Table 2.
The line fluxes of the prominent emission lines for all the observed epochs are given in Table 3. The striking feature of the $K$ band
spectra (Figure 6) is the presence of first overtone CO absorption bands with bandheads at 2.2935 $\mu$m and
beyond. They are direct confirmation that the system contains a cool secondary star. Similar CO features
are seen in a few other symbiotic systems such as RS Oph (Pavlenko et al 2008, Banerjee et al 2009),
V407 Cyg (Munari et al. 2011) and V745 Sco (Banerjee et al. 2014)- all of which contain late M type giants or
Mira variables as secondaries. Known cases of nova eruptions in
such symbiotic systems are relatively rare compared to eruptions in the standard configuration for
classical novae (i.e. outburst occurring in a system comprising of a WD + main sequence companion).
The present detection thus has an extra edge of interest attached to it.

\begin{figure}
\centering
\includegraphics[bb= 121 89 490 702,width=3.5in,height=5.0in,clip]{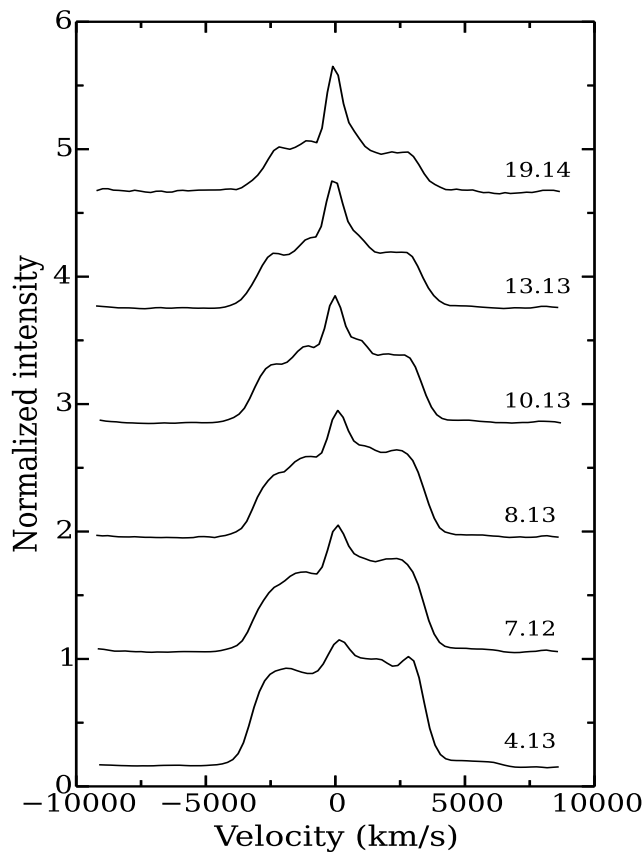}
\caption[]{ Evolution of the He\,{\sc i} 1.0831 $\mu$m line profile with time (see section 3.6)}
\label{fig3}
\end{figure}

\begin{figure}
\centering
\includegraphics[bb=121 89 490 702,width=3.5in,height=5.0in,clip]{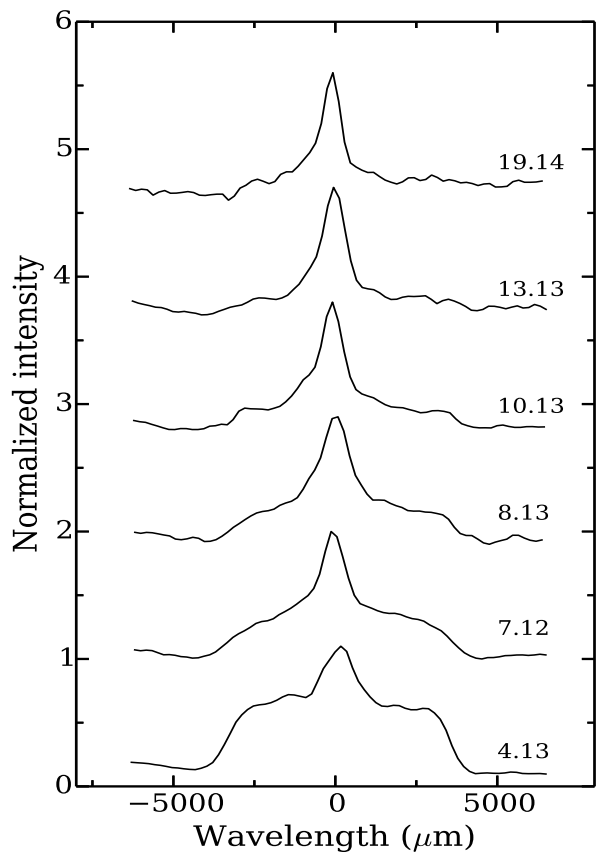}
\caption[]{ Evolution of the Pa$\beta$ 1.2818 $\mu$m line profile with time (see section 3.6)}
\label{fig3}
\end{figure}

\subsection{Evolution of line profiles}
We looked for the signature of a shock driven by the passage of the ejecta into the ambient
red giant wind. In three similar cases earlier viz RS Oph (Das et al. 2006), V407 Cyg (Munari et al. 2011, Banerjee et al. 2014)
and V745 Sco (Banerjee et al. 2014), evidence of a propagating shock was seen via the rapid narrowing of the NIR emission lines
as the shock front decelerated. The behavior of the shockwave as it propagates into the dense ambient medium surrounding the WD
consists of a free expansion or ejecta-dominated stage,
where the ejecta expands freely and the shock moves at a
constant speed without being slowed down by the surrounding
medium. This phase generally extends to the time it takes for the
 swept-up mass to equal the ejecta mass. The second phase is a Sedov-Taylor stage,
where the majority of the ejecta kinetic energy has been transferred
to the swept-up ambient gas. This is an adiabatic phase since the shocked material is so hot that
there is negligible cooling by radiation losses. During this phase a deceleration is seen in
the shock whose velocity decreases with time as a power law (more details in Das et al. 2006).
In the present case we do not see any sign of this deceleration.

The profiles of the He\,{\sc i} and Pa$\beta$ lines are shown in Figures 7 and 8 respectively.
The observed profiles are composed of a broad pedestal due to the high velocity nova ejecta
 on which is superposed a sharp and narrow component. Qualitatively, the profiles are very similar to those
 seen in V407 Cyg and V745 Sco in which the narrow component was thought to arise
 from a large fraction of the secondary's wind ionized by
the flash of energetic radiation produced by the thermonuclear event (Munari et al. 2010). However higher resolution data,
which should show the structures in the profiles in much greater detail, should be awaited to pinpoint the origin of the narrow component.
From Figures 7 and 8 visual inspection shows that the intensity of the broad component rapidly decreases with time as the
nova ejecta dilutes and fades. However its width remains fairly constant - this is more clearly seen for the stronger
He\,{\sc i} 1.0831 $\mu$m line. The estimated FWZI's for this line on the 6 epochs shown in Figure 7, in increasing order of
time, are 10100, 10200, 9900, 10100, 10200 and 10000 km s$^{-1}$ with a typical error of $\pm$ 25 km s$^{-1}$.
That is, the FWZI's remain constant and there is hence no evidence of a deceleration. This {\bf is} a puzzling feature for
which a straightforward or convincing explanation is not easily obtained.
 A possible explanation for this is that the ambient circum-binary medium is not dense enough to slow down the ejecta.
 Hence although a shock exists - albeit weak possibly - over the duration of our observations the ejecta remains in a
 free-expansion stage and does not enter the Sedov-Taylor decelerative phase. Otherwise it becomes difficult to reconcile
 the observed NIR behavior with the X-ray observations. These indicated the presence, almost immediately after the outburst began,
 of hot shocked gas with a modeled
 temperature of $kT$ = 6.4$^{+3.8}_{-2.1}$ keV i.e. close to 7$\times$10$^{7}$K (Kuulkers et al 2014; Page et al. 2014).

No detection of gamma-ray emission was reported by Fermi for this nova. When an energetic shock front exits, such emission can
be generated from diffusive acceleration of particles across the front.
 $\gamma$-ray detections, all emerging from Fermi LAT, are known from only six novae viz V407 Cyg, Nova Sco 2012, Nova Mon 2012,
 Nova Del 2013, Nova Cen 2013 and V745 Sco. Two of these (V407 Cyg and V745 Sco) are from symbiotic systems similar to the present
 nova. However the clearest detections at the best S/N, have been made for the nearby novae, the farthest of which is at 4.5 kpc
 (Ackermann et al. 2014). For a distant object e.g. in the case of V745 Sco, which is relatively farther away at an estimated
 distance of 7.8 $\pm$ 1.8 kpc (Schaefer 2010), the detection was weak at around the 2-3 sigma level (Cheung et al. 2014).
 So it is possible that $\gamma$-ray emission from this nova, if generated at all, could have been below or at the fringe of the Fermi detection limit.

\section{Discussion: The recurrent nova possibility}

The similarity of the secondary component in Nova Sco 2014 with the M giant secondaries seen in T CrB subclass of recurrent novae
(RNe) raises the possibility of Nova Sco 2014 being a RN (confirmed members of this subclass are T CrB, RS Oph, V3890 Sgr and V745 Sco).
Among the ten known recurrent novae, at least eight have evolved secondary stars unlike the main-sequence secondaries typical in classical
novae (Darnley et.al. 2012). Pagnotta $\&$ Schaefer (2014) have done an extensive study of a large sample of 237 known classical novae and
10 recurrent novae to identify RNe candidates among the known classical novae. They identify the characteristics that are common to RNe and
the available pre- and post-outburst data of
Nova Sco 2014 fulfils most of these characteristics, namely, small outburst amplitude, IR colors resembling
the colors of M type giants, expansion velocity exceeding 2000 km s$^{-1}$ and presence of high excitation lines
near the peak. It is also interesting to note that no CN has been found that has a M type giant secondary (Warner 1995).

The nature of the outburst of Nova Sco 2014 must
also be distinguished from those occurring in conventional symbiotic novae (sN). These eruptions, occuring
in symbiotic stars, comprise a total of about 10 objects including RR Tel, AG Peg, V1016 Cyg etc.
(Kenyon 1986). Their outburst characteristics however are very different from RN or CN.
Their eruptions are very protracted with the time between start of eruption and reaching maximum
running into years or decades. Similar time scales are observed for the brightness to drop by 1 magnitude.
The entire duration of the outburst also can run into tens of decades or even exceed a century. These aspects
are succinctly summarized in Warner (1995) and Kenyon (1986). Based on these arguments, the outburst
characteristics of Nova Sco 2014 suggest that there is a very strong possibility of it being a RN instead of a
CN; it is certainly very different from a sN. If it is indeed a RN, it is then desirable to study patrol plates of this
region to look for earlier outbursts.
Our search has been limited to OGLE data but the footprint of the OGLE galactic bulge survey,
ongoing for more than a decade, narrowly misses the region where the object lies. Preliminary search of the All Sky Automated
Survey (ASAS) data, between 2000 - 2009, is also unfruitful. However the ASAS limiting magnitude is close to 14 in $V$ and the
general cadence of observations is once every three days. Since this object peaks at $V$ = 12 and declines to 14 magnitude
in 6d, it could be missed by ASAS. We intend to make a more thorough and formal search for earlier outbursts from other archival data.

\section{Acknowledgments}
 We thank the anonymous referee for helpful comments that improved the paper. The research work at the Physical Research Laboratory is funded by
the Department of Space, Government of India. We acknowledge use of American Association of Variable Star Observers(AAVSO),
Two Micron All Sky Survey (2MASS) and Wide-field Infrared Survey Explorer (WISE) data.

\end{document}